\documentstyle[manuscript,aps]{revtex}
\begin{document}
\draft
\title{Integrable multiparametric quantum spin chains}
\author{Angela Foerster}
\address{Instituto de F\'{\i}sica da UFRGS \\[-1mm] Av. Bento Gon\c{c}alves 
9500,
Porto Alegre, RS - Brazil \\[-2mm] e-mail angela@if.ufrgs.br}
\author{Jon Links}
\address{Department of Mathematics \\[-1mm] University of
Queensland, Queensland, 4072,  Australia \\[-2mm]
e-mail jrl@maths.uq.oz.au}
\author{Itzhak Roditi \footnote{Permanent address: Centro Brasileiro de 
Pesquisas Fisicas - CBPF \\[-2mm]
Rua Dr. Xavier Sigaud 150, 22290-180, Rio de Janeiro, RJ - Brazil}}
\address{Theory Division, CERN \\[-1mm]
CH-1211 Geneva 23, Switzerland\\[-2mm]
e-mail Itzhak.Roditi@cern.ch}
\maketitle
\begin{abstract}
Using Reshetikhin's construction for multiparametric quantum algebras
we obtain the associated multiparametric quantum spin chains. We show
that under certain restrictions these models can be mapped
to quantum spin chains with twisted boundary conditions.
We illustrate how this general formalism applies to construct
multiparametric versions of the supersymmetric t-J and U models. \\
PACS: 71.20.Ad, 75.10.Jm, 71.27.+a \\
Keywords: Exactly solved models, Yang-Baxter algebra, quantum superalgebras,
systems of correlated electrons
\end{abstract}


\vspace{0.55cm}
\begin{flushleft}
CERN-TH/97-75 
\end{flushleft}

\vspace{0.55cm}
\begin{flushleft}
Corresponding author:
{\bf Dr. Jon Links} \\
Address: Department of Mathematics, \\
$\phantom{00000000}$ University of Queensland, \\
$\phantom{00000000}$ Queensland, 4072, Australia \\
e-mail: jrl@maths.uq.oz.au \\
Telephon: 61 7 336 53277 \\
Fax: 61 7 3365 1477
\end{flushleft}
\newpage
\section{Introduction }
The advent of quantum algebras \cite{Ji,Dr} precipitated many new
results in the area of integrable models. Likewise their supersymmetric
counterparts quantum superalgebras \cite{Zhan,y,kt,sch} facilitate systematic
treatments of integrable models which accommodate both bosonic and
fermionic degrees of freedom. An important subclass are those which may
be interpreted as describing systems of correlated electrons for their
obvious physical applications in condensed matter physics. Among these
are the supersymmetric (SUSY) $t-J$ model\cite{azr,ffk} and supersymmetric
generalizations of the Hubbard model \cite{eks,bglz}. Particularly, it would be
beneficial if these models could provide some knowledge of the
transitions between the metallic, insulating and superconducting phases.
It has long been known that an insight into these properties may be
gained by studying the effects of the boundary conditions on such
models (e.g. see \cite{by,ko}). 
Subsequently several authors have studied electronic
models with twisted boundary conditions \cite{ss,bares,aag,essler}.

Integrable models with twisted boundary conditions may be formulated
within the framework of quantum (super)algebras. This notion of twisting
is more general than that usually treated in electronic models whereby
a twisted boundary condition is thought of as the introduction of a
phase factor into the periodicity of the model. Here our twisted
boundary conditions correspond to more general transformations of the
local states and have their origins in the underlying symmetry of the
model \cite{dV}. For a given $R$-matrix $R(u)$ any matrix $M$ satisfying
$[M_1M_2,\,R(u)]=0$ allows an integrable model to be constructed with
the  trace over the auxiliary space weighted by $M$. This is what we
will refer to as a twisted boundary condition.

The work of Reshetikhin \cite{Re} on multiparametric quantum
(super)algebras permits a natural way to construct integrable models
also dependent upon additional free parameters. In fact the model of
Perk and Schultz \cite{PS} may be formulated within this framework
\cite{OY}. We will make it apparent
that there exists under suitable conditions a
integrability preserving mapping between these
multiparametric models and those with twisted boundary conditions.
Some particular cases have been studied previously \cite{abb,ps,rrrs,frr}.
We show that
these methods may be applied to any model with an underlying quantum
(super)algebra symmetry.  By establishing this correspondence between
models with twisted boundary conditions and multiparametric models, it
is reasonable to expect that the multiparametric solutions 
may provide suitable test models for
describing the various phases associated with correlated electron
systems.

As examples we will consider the cases of the SUSY $t-J$ model
\cite{ffk} and the SUSY $U$ model \cite{bglz}. 
As well as giving the Hamiltonians for the integrable multiparametric
generalizations of these two models, we have also determined the
corresponding Bethe ansatz equations which provide the starting point
for an investigation into their thermodynamics.

The paper is organized as follows. In section II we present the general
construction of multiparametric spin chains and their relation to
models with twisted boundary conditions. In section III we illustrate
how our general formalism applies to construct multiparametric versions
of the SUSY t-J and U models. The Bethe ansatz equations of the
models are also obtained. A summary of our main results is presented
in section IV.

\section{General Construction}
Using Reshetikhin's construction \cite{Re} for multiparametric
quantum
algebras, it is straightforward to obtain the associated
multiparametric
quantum spin chain. Here we  demonstrate  that under appropriate
constraints these models may be transformed to quantum spin chains 
with
twisted boundary conditions; i.e. the additional parameters arising
from
Reshetikhin's construction may be mapped to the boundary.

Let $(A,~\Delta,~R)$ denote a quasitriangular Hopf (super)algebra
where
$\Delta$ and $R$ denote the co-product and $R$-matrix respectively.
Suppose that there exists an element $F\in A\otimes A$ such that
\begin{eqnarray}
(\Delta \otimes I)(F)=&F_{13}F_{23},~~~~~~(I\otimes
\Delta)(F)&=F_{13}F_{12},   \nonumber    \\
F_{12}F_{13}F_{23}=&F_{23}F_{13}F_{12},~~~~~~F_{12}F_{21}&=I 
\label{fyb}
\end{eqnarray}
Theorem 1 of \cite{Re} states that $(A,~\Delta^{F},~R^F)$ is also a
quasitriangular Hopf algebra with co-product and $R$-matrix 
respectively
given by
\begin{equation}
\Delta^F=F_{12}\Delta F_{21},~~~~~~~~R^F=F_{21}RF_{21}.
\label{df}
\end{equation}
In the case that $(A,~\Delta,~R)$ is an affine quantum (super)algebra 
we
have from \cite{Re} that $F$ can be chosen to be
\begin{equation}
F={\rm exp} \sum_{i<j}\left(H_i\otimes H_j-H_j\otimes
H_i\right)\phi_{ij}
\label{F}  \end{equation}
where $\{H_i\}$ is a basis for the Cartan subalgebra of the affine
quantum (super)algebra and the $\phi_{ij},~i<j$ are arbitrary complex
parameters. For our purposes we will extend the Cartan subalgebra by 
an
additional central extension (not the usual central charge) $H_0$
which
will act as a scalar multiple of the identity operator in any 
representation.

Suppose that $\pi$ is a loop representation of the affine quantum
superalgebra. We let $R(u)$, $R^F(u)$ be the (super)matrix
representatives of $R$ and $R^F$ respectively, which both satisfy the
Yang-Baxter equation
$$R_{12}(u-v)R_{13}(u)R_{23}(v)=R_{23}(v)R_{13}(u)R_{12}(u-v).$$
For the supersymmetric case it is necessary to impose the multiplication
rule
$$(a\otimes b)(c\otimes d)=(-1)^{[b][c]}(ac\otimes bd)$$
for homogeneous supermatrices $a,\,b,\,c,\,d$ where $[a]=0$ if $a$ is even and
$[a]=1$ if $a$ is odd \cite{jsv}. However with an appropriate
redefinition of its
matrix elements $R(u)$ satisfies the usual (non-graded) Yang-Baxter
equation (e.g. see \cite{dglz}). Throughout we adopt this latter
convention.

If $\left.R(u)\right|_{u=0}=P$ with $P$ the permutation operator then
$\left.R^F(u)\right|_{u=0}=P$
as a result of (\ref{fyb}). We may construct the transfer matrix
\begin{eqnarray}
t^F(u)&=&{\rm str}_0\left(\pi^{\otimes(N+1)} 
\left(I\otimes\Delta^F_N\right)R^F_{01}\right)  \nonumber
\\
&=&{\rm str}_0\left(R^F_{0N}(u)R^F_{0(N-1)}(u)....R^F_{01}(u)
\right)
\label{tF}
\end{eqnarray}
where $\Delta^F_N$ is defined recursively through
\begin{eqnarray}
\Delta^F_N&=&\left(I\otimes I....\otimes 
\Delta^F\right)\Delta^F_{N-1}
\nonumber   \\
&=&\left(\Delta^F\otimes I....\otimes I\right)\Delta^F_{N-1}
.\end{eqnarray}
The subscripts 0 and 1,2,...,$N$ denote the auxiliary and quantum spaces
respectively and ${\rm str}_0$ is the supertrace over the zeroth space.
From the Yang-Baxter equation it follows that the multiparametric
transfer matrices $t^F(u)$ form a commuting family.
The associated multiparametric spin chain Hamiltonian is given by
\begin{eqnarray}
H^F&=&\left.\left(t^F(u)\right)^{-1}\frac{d}{du}t^F(u)\right|_{u=0}  
\nonumber  \\
&=&\sum_{i=1}^{N-1}h^F_{i,i+1}+h^F_{N1}   \label{HF} \end{eqnarray}
 with
 $$h^F=\frac{d}{du}\left.PR^F(u)\right|_{u=0}.  $$

Through use of (\ref{fyb}) we may alternatively write
$$t^F(u)={\rm str}_0 \left(\pi^{\otimes(N+1)}\left(I\otimes 
J_N\right)
\left[(I\otimes
\Delta_N)(F_{10}R_{01}F_{10})\right]\left(I\otimes
J_N\right)^{-1}\right)$$
with
\begin{eqnarray}
J_N&=&G_{N-1}G_{N-2}....G_1,   \nonumber  \\
G_i&=&F_{iN}F_{i(N-1)}....F_{i(i+1)}.    \end{eqnarray}
We now define a new transfer matrix
\begin{eqnarray}
t(u)&=& J_N^{-1}t^F(u) J_N  \nonumber \\
&=&{\rm str}_0\left(\pi^{\otimes(N+1)}\left(I\otimes\Delta_N\right)
\left(F_{10}R_{01}F_{10}\right)\right) \label{77}
\end{eqnarray}
where we have employed the convention to let $F$ denote both the
algebraic object and its (super)matrix representative. Through 
further
use of (\ref{fyb}) we may show that
$$t(u)={\rm str}_0\left(F_{10}F_{20}....F_{N0}R_{0N}(u)R_{0(N-1)}(u)
....R_{01}(u)F_{10}....F_{N0}\right)$$
and the associated Hamiltonian is given by
\begin{eqnarray}
H&=&\left.\left(t(u)\right)^{-1}\frac{d}{du} t(u)\right|_{u=0}  
\nonumber  \\
&=&\sum_{i=1}^{N-1}h_{i,i+1} + \left(F_{N(N-1)}....F_{N1}\right)^2
h_{N1} \left(F_{1N}....F_{(N-1)N}\right)^2   \label{ham}
\end{eqnarray}
where
$$h=\left.\frac{d}{du}PR(u)\right|_{u=0}.  $$

The above Hamiltonian describes a closed system where instead of the
usual periodic boundary conditions we now have a more general type of
boundary condition. The boundary term in the above Hamiltonian is a
global operator; i.e. it acts non-trivially on all sites. However we can
in fact interpret this term as a local operator which couples only the
sites labelled 1 and $N$. It can be shown that the boundary term 
commutes with the local observables $h_{i,i+1}$ for $i\neq
1,~N-1$.
This situation is analogous to the closed quantum (super)algebra
invariant chains discussed in \cite{lf}.

From the above construction we may also yield models with twisted
boundary conditions by an appropriate choice of $F$. Recall that we
extend the Cartan subalgebra by the central element $H_0$. Let this
element act as $cI$ in the representation $\pi$ where $c$ is some
complex number. If we now choose $\phi_{ij}=0$ for $i\neq 0$ in the
expression (\ref{F}) the matrix $F$ factorizes as $F=M_1^{-1}M_2$
with
$$M={\rm exp}\left(\sum_{i=1}^l c\phi_{0i}H_i\right)      $$
and $l$ is the rank of the underlying quantum (super)algebra
$U_q(g)$.
Using the fact that the $R$-matrix satisfies
$$\left[R(u),~I\otimes H_i+H_i\otimes I\right]=0,\qquad i=1,2,...,l$$
 tells us that
$$\left[R(u),~M_1M_2\right]=0.$$
In this case the Hamiltonian (\ref{ham}) reduces to
\begin{equation}
H = \sum_{i=1}^{N-1} h_{i,i+1}+ M_1^{2N}h_{N1}M_1^{-2N}
\label{htbc}
\end{equation}
which is precisely the form for a system with twisted boundary
conditions (see \cite{dV}).

\section{Examples}
In this section we illustrate how our formalism applies to construct
a multiparametric version of the SUSY
t-J model \cite{azr} and the SUSY U model \cite{bglz}. Both models
are $gl(2/1)$ invariant and their formulation through the
quantum inverse scattering method can be found in
\cite{ffk} and \cite{km}, respectively. The first model describes
electrons with nearest-neighbor hopping and spin exchange
interaction on a chain, while the second can be considered
an extension of the Hubbard model with additional pair-hopping
and bond-charge interaction terms. These models are of interest
because of their possible connection with high-Tc superconductivity.
In order to turn our discussion more general, we will in fact handle with
their anisotropic or q-deformed versions \cite{FK3}, \cite{bkz,ghlz}.
Of course, in the rational limit $q \rightarrow 1$ all results
reduce to their corresponding isotropic ones.

\subsection{The supersymmetric t-J model}
We begin by introducing the multiparametric 
$U_q(g\ell(2/1))$ $ R $-matrix, which in terms of a
generic spectral parameter $x$ and a deformation parameter $q$ reads
\begin{equation}
{\footnotesize
R^F(x)=
\pmatrix{
 \ a &0 &0 &0 &0  &0  &0  &0  &0  \cr
  0 & t^2_1 b  &0 &c_{-} &0  &0  &0  &0  &0 \cr
   0 &0 & t^2_2 b &0 &0  &0  &c_{-} &0 &0 \cr
    0 &c_{+} &0 &\frac{b}{t^2_1}  &0  &0  &0  &0  &0 \cr
     0 &0 &0 &0 &a &0  &0  &0  &0 \cr
      0 &0 &0 &0 &0  &t^2_3 b  &0  &c_{-}  &0 \cr
       0 &0 &c_{+} &0 &0  &0  &\frac{b}{t^2_2}  &0  &0 \cr
	0 &0 &0 &0 &0  &c_{+}  &0  &\frac{b}{t^2_3}  &0 \cr
	 0 &0 &0 &0 &0  &0  &0  &0 &w }}
	 \label{rftj}
	 \end{equation}
where
\begin{equation}
a = x q - { 1 \over x q } ,\quad
b = x - { 1 \over x } ,\quad
c_+ = x ( q - { 1 \over q} ), \quad
c_- = { 1 \over x } (q - { 1 \over q} ),\quad
w = -{ x \over q}  + { q \over x }\, \, \, 
\end{equation}
and $t_1, t_2, t_3$ are independent parameters written in terms of
$\phi_{01}, \phi_{02}, \phi_{12}$ as
\begin{eqnarray}
t_1 &=& {\rm exp} (- 2 \phi_{01} + \phi_{02} + \phi_{12} ) \nonumber \\
t_2 &=& {\rm exp} (- \phi_{01} + \phi_{02} + \phi_{12} ) \label{t1t2t3} \\
t_3 &=& {\rm exp} ( \phi_{01} - \phi_{12} ) \nonumber
\end{eqnarray}

The above matrix was already presented by Perk and
Schultz \cite{PS} when studying a multicomponent
generalization of the six-vertex model.

Next we construct the transfer matrix $t^F (x)$ according to
eq.(\ref{tF}) from which we find the
associated multiparametric Hamiltonian (see eq.(\ref{HF}) )
on a one-dimensional periodic lattice
$$H^F=\sum_{i=1}^{N-1}h^F_{i,i+1}+h^F_{N1}, $$
where
\begin{eqnarray}
h^F_{i,i+1}&=& - (
\frac{1}{t^2_2}  c^{\dagger}_{i\uparrow} c_{i+1\uparrow} +
t^2_2  c^{\dagger}_{i+1\uparrow} c_{i\uparrow} +
\frac{1}{t^2_3}  c^{\dagger}_{i\downarrow} c_{i+1\downarrow} +
t^2_3  c^{\dagger}_{i+1\downarrow} c_{i\downarrow} ) \nonumber \\
 \nonumber
 &-& 2 \biggl[ \frac{1}{t^2_1}  S_i^+ S_{i+1}^- +
 t^2_1 S_i^- S_{i+1}^ +
 + 2 \cos \gamma \phantom{0}(S_i^z
 S_{i+1}^z - {n_i n_{i+1}\over 4}) \biggr]  \\
 &+& i\sin( \gamma ) (n_i - n_{i+1})
 -i\sin ( \gamma)(n_i S_{i+1}^z - S_i^z n_{i+1})
 - \cos \gamma n_i + 2 \cos \gamma
\label{hftj}
\end{eqnarray}
Above $c_{i\pm}^{(\dagger)}$'s are spin up or down annihilation
(creation) operators, the $\vec S_i$'s
spin matrices, the $n_i$'s occupation numbers of electrons at
lattice site $i$ and $\gamma$ is the anisotropy parameter
($q = e^{i \gamma} $). A similar version of a multiparametric
SUSY
t-J model has already been discussed in ref. \cite{Ar}. Notice
that here it emerges systematically from our general construction.
By setting $t_1, t_2, t_3 \rightarrow 1$ in eq. (\ref{hftj}),
the usual terms of the
anisotropic SUSY t-J model \cite{FK3} can be recovered.

The Hamiltonian  (\ref{hftj})
can be exactly solved through the algebraic nested Bethe ansatz
method.
This procedure is carried out in two steps and the Bethe ansatz
equations
are given by
\begin{eqnarray}
&&t^{2(N-M_2)}_1 t^{2 M_2}_2 t^{-2 M_2}_3  {\biggl(
{a(x^{(1)}_k) \over b(x^{(1)}_k) }  \biggr)}^N
\prod_{i=1}^{M_1} { a(x^{(1)}_i / x^{(1)}_k) \over b(x^{(1)}_i /
x^{(1)}_k)}
{ b(x^{(1)}_k / x^{(1)}_i) \over a(x^{(1)}_k / x^{(1)}_i)}
\prod_{j=1}^{M_2} { b(x^{(2)}_j / x^{(1)}_k) \over a(x^{(2)}_j /
x^{(1)}_k)}
= -1 \, \,  , \, \,
k = 1, \dots M_1  , \nonumber \\
&&t^{-2(N-M_1)}_1 t^{2(N-M_1)}_2 t^{2 M_1}_3 {(-1)}^{M_2} \prod_{i=1}^{M_1}
{ a(x^{(2)}_k / x^{(1)}_i) \over b(x^{(2)}_k / x^{(1)}_i)}
\prod_{j=1}^{M_2} { a(x^{(2)}_j / x^{(2)}_k) \over b(x^{(2)}_j /
x^{(2)}_k)}
{ b(x^{(2)}_k / x^{(2)}_j) \over w(x^{(2)}_k / x^{(2)}_j)}
= 1 \, \, ,  \, 
\,
k = 1, \dots M_2
\label{baetj}
\end{eqnarray}
where $ x^{(m)}_k (m=1,2~;~k=1,...,M_m)$ denote the Bethe ansatz
parameters,
$N$ is the number of lattice sites, $M_1$ is the number of holes plus
down spins and $M_2$ is the number of holes. We see from
(\ref{baetj}) that the additional parameters $t_1, t_2, t_3$ have
the meaning of external fields (see e.g. \cite{vega,Hl})

Following the approach presented in the previous section,
we perform the transformation  (\ref{77}) and then set
$t_3=\frac{t_2}{t_1}$ (or $\phi_{12} = 0$ , see eq. (\ref{t1t2t3}))
in order to find the anisotropic SUSY t-J model
with twisted boundary conditions (\ref{htbc})
$$H=\sum_{i=1}^{N-1}h_{i,i+1}+h_{N1}, $$
where $h_{i,i+1} = \lim_{ \{t_1, t_2, t_3
\rightarrow 1 \} } h^F_{i, i+1}$ and
\begin{eqnarray}
h_{N,1}&=& - \biggl[
t^{2N}_2 c^{\dagger}_{N\uparrow} c_{1\uparrow} +
\frac{1}{t^{2N}_2} c^{\dagger}_{1\uparrow} c_{N\uparrow} +
{ \biggl( \frac{t_2}{t_1} \biggr) }^{2N}  c^{\dagger}_{N\downarrow}
c_{1\downarrow} +
{ \biggl( \frac{t_1}{t_2} \biggr) }^{2N}  c^{\dagger}_{1\downarrow}
c_{N\downarrow} \biggr] \nonumber \\
\nonumber
&-& 2 \biggl[ t^{2N}_1 S_N^+ S_{1}^- +
\frac{1}{t^{2N}_1} S_N^- S_{1}^ +
+2 \cos \gamma \phantom{0}(S_N^z
S_{1}^z - {n_N n_{1}\over 4}) \biggr]  \\
&+& i\sin( \gamma ) (n_N - n_{1})
-i\sin ( \gamma)(n_N S_{1}^z - S_N^z n_{1})
- \cos \gamma n_N + 2 \cos \gamma
\label{hftjtb}
\end{eqnarray}
\subsection{The supersymmetric U model}
Let us now construct a
multiparametric version of the anisotropic SUSY U model,
which has been proposed  recently as a new integrable model for
correlated electrons (see  ref. \cite{bkz,ghlz}  for more details).

We begin by recalling the trigonometric R-matrix associated with the
one
parameter family of four-dimensional representations of
$U_q(g\ell(2/1))$
\begin{eqnarray}
R(x)&=&P\check{R}(x), \nonumber \\
\check{R}(x)&=& \frac {q^x - q^{2 \alpha}} {1- q^{x + 2 \alpha}} P_1
+ P_2
+\frac {1 - q^{x+2 \alpha +2}} {q^x- q^{2 \alpha + 2}} P_3.
\label{rum}
\end{eqnarray}
Here $x$ and $q$ are, respectively, the spectral and deformation
parameters and $\alpha$ is a free parameter which arises from the
underlying  representation. $P$ is the permutation operator and
$P_i, i = 1,2,3$ are projectors whose explicit form can be found in
\cite{ghlz}.

We find the corresponding multiparametric $R$-matrix
%
%
\begin{equation}
{\footnotesize
R^F(x)=
\pmatrix{
 * &0 &0 &0 &0  &0  &0  &0  &0  &0  &0  &0  &0  &0  &0  &0
\cr
 0 & t^2_1 * &0 &0 & *  &0  &0  &0  &0  &0  &0  &0  &0  &0  &0
&0    \cr
 0 &0 &t^2_2 * &0 &0  &0  &0  &0  & * &0  &0  &0  &0  &0  &0  &0
  \cr
 0 &0 &0 &t^2_1 t^2_2 * &0  &0  & t_1 t_2 t_3 * &0  &0
&\frac{t_1 t_2}{t_3}*  &0  &0  &*  &0  &0  &0    \cr
 0 &* &0 &0 &\frac{*}{t^2_1} &0  &0  &0  &0  &0  &0  &0  &0  &0
&0  &0    \cr
 0 &0 &0 &0 &0  &*  &0  &0  &0  &0  &0  &0  &0  &0  &0  &0    \cr
 0 &0 &0 &t_1 t_2 t_3 * &0  &0  &t^2_3 *  &0  &0  & * &0  &0  &
\frac{t_3}{t_1 t_2}* &0  &0  &0    \cr
 0 &0 &0 &0 &0  &0  &0  &t^2_1 t^2_3 *  &0  &0  &0  &0  &0  &*
 &0  &0    \cr
 0 &0 &* &0 &0  &0  &0  &0  &\frac{*}{t^2_2}  &0  &0  &0  &0  &0
&0  &0    \cr
 0 &0 &0 & \frac{t_1 t_2}{t_3}*  &0  &0  & * &0  &0  &
\frac{*}{t^2_3} &0  &0  & \frac{*}{t_1 t_2 t_3}  &0  &0  &0    \cr
 0 &0 &0 &0 &0  &0  &0  &0  &0  &0  &*  &0  &0  &0  &0  &0    \cr
 0 &0 &0 &0 &0  &0  &0  &0  &0  &0  &0  &  * \frac{t^2_2}{t^2_3}
 &0  &0  & *  &0    \cr
 0 &0 &0 & * &0  &0  & \frac{t_3}{t_1 t_2} *  &0  &0
& \frac{*}{t_1 t_2 t_3}  &0  &0  &\frac{*}{t^2_1 t^2_2}  &0  &0  &0    \cr
 0 &0 &0 &0 &0  &0  &0  &* &0  &0  &0  &0  &0  &
\frac{*}{t^2_1 t^2_3}&0  &0    \cr
 0 &0 &0 &0 &0  &0  &0  &0  &0  &0  &0  &*  &0  &0  &  *
\frac{t^2_3}{t^2_2} &0    \cr
 0 &0 &0 &0 &0  &0  &0  &0  &0  &0  &0  &0  &0  &0  &0  & *}
}
\end{equation}
%
%
%
%
where $t_1$, $t_2$ and $t_3$ are independent parameters also given by
eq. (\ref{t1t2t3}) and  ``$*$'' denote the elements of the $R$-matrix
(\ref{rum}), which can be obtained from the projectors given in 
\cite{ghlz}. We do not write them explicitly here since we will
not need them later.  Notice that here, in contrast to the previous
case (see eq.(\ref{rftj})), the new parameters $t_1, t_2, t_3$
occupy also non-diagonal
positions. This is a peculiarity of higher representations and can
also be verified for other higher spin models ( e. g. spin 1 XXZ
chain). In fact, our prescription for the element $F$ of the
multiparametric $R$ matrix is particularly interesting in these
cases, where it is not obvious how to construct $R^F$.

Next we construct the transfer matrix $t^F (x)$ according to
eq.(\ref{tF}) from which we find the multiparametric
version of the anisotropic SUSY U model (see eq.(\ref{HF})
)
$$H^F=\sum_{i=1}^{N-1}h^F_{i,i+1}+h^F_{N1},$$ where
\begin{eqnarray}
h^F_{i,i+1}&=&-\xi c^+_{i\uparrow} c_{{i+1}\uparrow}
[-{\eta}^{-1}\nu^+]^{n_{i\downarrow}} 
[-{\eta}^{-1}\nu^-]^{n_{{i+1}\downarrow}} - h.c.  \nonumber\\
           & &- \rho c^+_{i\downarrow} c_{{i+1}\downarrow}  
[-{\eta} \nu^-]^{n_{i\uparrow}} 
[-{\eta} \nu^+]^{n_{{i+1}\uparrow}} 
 - h.c.\nonumber\\
           & &+ [\alpha]^{-1}_q \xi \rho c^+_{i\downarrow}
c^+_{i\uparrow} c_{{i+1}\uparrow} c_{{i+1}\downarrow} + h.c.
\nonumber\\    
           & & + [\alpha]^{-1}_q (n_{i\uparrow} n_{i\downarrow}  +  
n_{{i+1}\uparrow} n_{{i+1}\downarrow})\nonumber \\
           & &+ q^{\alpha + 1} (n_{i\uparrow} + n_{i\downarrow} -1) +
q^{-\alpha - 1} (n_{{i+1}\uparrow} + n_{{i+1}\downarrow} -1)
\label{hfii} \end{eqnarray}
and $\xi=\frac {t^2_3}{t^2_2}$ ,  $\eta=\frac {t_1 t_3}{t_2}$ ,
$\rho=\frac {1}{t^2_1 t^2_3}$ , $\nu^{\pm}=sgn(\alpha)
q^{\pm \frac{1}{2}}(\frac{[\alpha + 1]_q}{[\alpha]_q})^{\frac{1}{2}}$.
In the $ h.c.$ terms one should notice that the parameters
$t_i \rightarrow t_i^{-1}, i=1,2,3$.
Equation (\ref{hfii}) is a generalization of the
anisotropic SUSY
U model \cite{ghlz}, which can be recovered in the limit
$t_1, t_2, t_3 \rightarrow 1$ .

This model can be exactly solved by means of the algebraic
nested Bethe ansatz method and the Bethe ansatz equations are
given by
\begin{eqnarray}
t^{2(N-N_2)}_1 t^{2 N_2}_2 t^{-2 N_2}_3
\biggl( \frac { x^{(1)}_k q^{\alpha +1} - q^{-1} }
{ x^{(1)}_k q^{-\alpha} -1 } \biggr)^N &=& \prod_{j \ne k}^{N_2}
\frac { x^{(1)}_k - x^{(2)}_j q^2 }{q(x^{(1)}_k -
x^{(2)}_j)}~~~;~~~~~~k=1, \dots N_1 \nonumber \\
t^{-2(N-N_1)}_1 t^{2(N-N_1)}_2 t^{2 N_1}_3 \prod_{i}^{N_1} \frac { x^{(1)}_i
- x^{(2)}_k q^2 }{q(x^{(1)}_i - x^{(2)}_k)} &=& \prod_{j \ne k}^{N_2}
\frac { -x^{(2)}_j + x^{(2)}_k q^2 }{ x^{(2)}_k - x^{(2)}_j q^2
}~~~;~~~~~~k=1, \dots N_2
\label{BAE}
\end{eqnarray}
where $x^{(m)}_k (m=1,2~;~k=1,...,N_m)$ are the Bethe ansatz
parameters, $N_1$ is the total number of
spins and $N_2$ is the number of spins down.

According to the approach presented in the previous section,
we perform the transformation (\ref{77}) and then choose
$t_3=\frac{t_2}{t_1}$
in order to find the anisotropic SUSY U model
with twisted boundary conditions, which yields
$$H=\sum_{i=1}^{N-1}h_{i,i+1}+h_{N1}, $$
where $h_{i,i+1}$ denotes the local terms of the anisotropic
SUSY U model \cite{ghlz} and
\begin{eqnarray}
h_{N,1}&=& -t_1^{2N} c^+_{N\uparrow} c_{{1}\uparrow}
 (-\nu^+)^{n_{N\downarrow}}
 (-\nu^-)^{n_{{1}\downarrow}}
 -  h.c.  \nonumber\\
           & &- t_2^{2N} c^+_{N\downarrow} c_{{1}\downarrow}
 (-\nu^-)^{n_{N\uparrow}}
 (-\nu^+)^{n_{{1}\uparrow}}
 -  h.c.\nonumber\\
           & &+ [\alpha]^{-1}_q (t_1 t_2)^{-2N} ~ c^+_{N\downarrow}
c^+_{N\uparrow} c_{{1}\uparrow} c_{{1}\downarrow} +h.c.\nonumber \\
          & &+ [\alpha]^{-1}_q (n_{N\uparrow} n_{N\downarrow}  +
n_{{1}\uparrow} n_{{1}\downarrow})\nonumber \\
           & &+ q^{\alpha + 1} (n_{N\uparrow} + n_{N\downarrow} -1) +
q^{-\alpha - 1} (n_{{1}\uparrow} + n_{{1}\downarrow} -1)
\label{btu} \end{eqnarray}

\section{Conclusions}
In this paper we have demonstrated a correspondence between
multiparametric spin chains
and models with twisted boundary conditions in the expectation
that this connection will provide further insight into the description
of phase transitions of such integrable systems. Our approach can be
applied to any model with an underlying quantum (super)algebra symmetry.
We are particularly
interested in models which describe systems of correlated electrons and
have studied the SUSY $t-J$ and $U$ models as examples.

Another important class of integrable models are those associated with
the Temperley-Lieb algebra. In \cite{z} Zhang proposes a systematic
method to generate multiparametric extensions of these models. It is
possible to adapt the techniques employed in this paper to establish a
mapping from models based on the Temperley-Lieb algebra with
twisted boundary conditions to associated multiparametric
generalizations. With
respect to correlated electron systems, an example based on the
Temperley-Lieb algebra has been described in \cite{l,flr} to which this
procedure can be applied.

\vspace{0.55cm}

ACKNOWLEDGEMENTS

The authors wish to thank J.H.H. Perk for illuminating discussions held
at the International Workshop on Statistical Mechanics and Integrable
Systems at Coolangatta, Australia, July 1997. 
We also thank D. Arnaudon for helpful comments.
A.F. would like to thank CNPq (Conselho Nacional de Desenvolvimento
Cient\'{\i}fico e Tecnol\'ogico) for financial support.
J.L. is supported by an Australian Research Council Postdoctoral
Fellowship.
I.R. thanks CERN and the Instituto de F\'{\i}sica da UFRGS for
their hospitality and CNPq for financial support.

\newpage


%
%

%
%


\begin{references}
\bibitem{Ji} M.~Jimbo, Lett. Math. Phys. {\bf 10} (1985) 63
\bibitem{Dr} V.G.~Drinfeld, Proc. Int. Cong. Math., Berkeley (1986) 798
\bibitem{Zhan} A.J.~Bracken, M.D.~Gould and R.B.~Zhang, Mod. Phys.
Lett. {\bf A5} (1990) 831
\bibitem{y} H.~Yamane, Proc. Jpn. Acad. {\bf A67} (1991) 108
\bibitem{kt} S.M.~Khoroshkin and V.N. Tolstoy, Commun. Math. Phys. {\bf
141} (1991) 599
\bibitem{sch} M.~Scheunert, J. Math. Phys. {\bf 34} (1993) 3780
\bibitem{azr} P.W.~Anderson, Science {\bf 235} (1987) 1196  \\
F.C.~Zhang and T.M.~Rice, Phys. Rev. {\bf B37} (1988) 3759
\bibitem{ffk} F.H.L.~Essler and V.E.~Korepin, Phys. Rev. {\bf B46}
(1992) 9147; \\
A.~Foerster and M.~Karowski, Phys. Rev.
{\bf B46} (1992) 9234; Nucl. Phys. {\bf B396} (1993) 611
\bibitem{eks} F.H.L.~Essler, V.E.~Korepin and K.~Schoutens, Phys. Rev.
Lett. {\bf 68} (1992) 2960; ibid. {\bf 70} (1993) 73
\bibitem{bglz} A.J.~Bracken, M.D.~Gould, J.R.~Links and Y.-Z.~Zhang,
Phys. Rev. Lett. {\bf 74} (1995) 2769
\bibitem{by} N.~Byers and C.N.~Yang, Phys. Rev. Lett. {\bf 7} (1961) 46
\bibitem{ko} W.~Kohn, Phys. Rev. {\bf 133} (1964) A171
\bibitem{yang} C.N.~Yang, Rev. Mod. Phys. {\bf 34} (1962) 696
\bibitem{yang1} C.N.~Yang, Phys. Rev. Lett. {\bf 63} (1989) 2144
\bibitem{yz} C.N.~Yang and S.~Zhang, Mod. Phys. Lett. {\bf B4} (1990)
759
\bibitem{ss} B.S.~Shastry and B.~Sutherland, Phys. Rev. Lett. {\bf 65}
(1990) 243
\bibitem{bares} P.A.~Bares, J.M.P.~Carmelo, J.~Ferrer and P.~Horsch,
Phys. Rev. {\bf B46} (1992) 14624 
\bibitem{aag} L.~Arrachea, A.A.~Aligia and E. Gagliano, Phys. Rev. Lett.
{\bf 76} (1996) 4396
\bibitem{essler} G.~Bed\"urftig, F.H.L.~Essler and H.~Frahm, Nucl. Phys
{\bf B498} (1997) 697
\bibitem{dV} H.J.~de Vega, Nucl. Phys. {\bf B240} (1984) 495; \\
C.M.~Yung and M.T.~Batchelor, Nucl. Phys. {\bf446} [FS] (1995) 461
\bibitem{Re} N.~Reshetikhin, Lett. Math. Phys. {\bf 20} (1990) 331
\bibitem{PS} J.H.H.~Perk and C.L.~Schultz, Phys. Lett. {\bf 84A} (1981)
407;
C.L.~Schultz, Phyisca {\bf 122A} (1983) 71
\bibitem{OY} M.~Okado and H.~Yamane, ICM-90 Satellite Conference
Proceedings, Eds. M.~Kashiwara and T.~Miwa (Springer-Verlag, 1990) 289
\bibitem{abb} F.C.~Alcaraz, M.N.~Barber and M.T. Batchelor, Annals of
Phys. {\bf 182} (1988) 280
\bibitem{ps} V.~Pasquier and H.~Saleur, Nucl. Phys. {\bf B316} (1990) 523
\bibitem{rrrs} M.~R-Monteiro, I.~Roditi, L.M.C.S.~Rodrigues, S.~Sciuto
Phys. Lett. {\bf B354} (1995) 389
\bibitem{frr} A.~Foerster, I.~Roditi and L.M.C.S.~Rodrigues, Mod. Phys.
Lett. {\bf A11} (1996) 987
\bibitem{jsv} P.P.~Kulish and E.K.~Sklyanin, J. Sov. Maths. {\bf 19}
(1982) 1596
\bibitem{dglz} G.W.~Delius, M.D.~Gould, J.R.~Links and Y.-Z.~Zhang, Int.
J. Mod. Phys. {\bf A10} (1995) 3259.
\bibitem{lf} J.~Links and A.~Foerster, J. Phys. {\bf A30} (1997) 2483
\bibitem{km} K.E.~Hibberd, M.D.~Gould, J.R.~Links, Phys. Rev. {\bf B54} (1996) 
8430; \\
P.B.~Ramos and M.J.~Martins, Nucl. Phys. {\bf B474} [FS] (1996) 678
\bibitem{FK3} A.~Foerster and M.~Karowski, Nucl. Phys. B
{\bf B408}, (1993) 512
\bibitem{bkz} R.Z.~Bariev, A. Kl\"umper and J.~Zittartz, Europhys.
Lett.  {\bf 32} (1995) 85 
\bibitem{ghlz} M.D.~Gould, K.E.~Hibberd, J.R.~Links and Y.-Z.~Zhang,
Phys. Lett. {\bf A212} (1996) 156 \\
K.E.~Hibberd, M.D.~Gould, J.R.~Links, J. Phys. {\bf A29} (1996) 8053
\bibitem{Ar} D.~Arnaudon, C. Chryssomalakos and L. Frappat, J. Math. Phys. {\bf
36} (1995) 5262 
\bibitem{vega} H.J.~de Vega and E.~Lopes, Phys. Rev. Lett. {\bf 67} (1991)
489
\bibitem{Hl} L. Hlavat\'y, Solution to the YBE corresponding to the
XXZ models in an external magnetic field, {\it Preprint E5-85-959,
Dubna} (1985)
\bibitem{z} R.B.~Zhang, J. Phys. {\bf A24} (1991) L535
\bibitem{l} J.~Links, J. Phys. {\bf A29} (1996) L69
\bibitem{flr} A.~Foerster, J.~Links and I.~Roditi, Mod. Phys. Lett.
{\bf A12} (1997) 1035 
\end{references}
\end{document}